\documentclass[aps,prb,twocolumn,oneside,floatfix,sort&compress,
superscriptaddress,showpacs,balancelastpage]{revtex4-1}

\usepackage{graphicx,xcolor}
\usepackage[utf8]{inputenc}
\usepackage{amssymb,bbm}
\usepackage[reqno]{amsmath}

\usepackage[colorlinks,bookmarks=false,citecolor=blue,linkcolor=red,urlcolor=blue,hyperfootnotes=true]{hyperref}

\begin{document}

\title{Optical conductivity of the Hubbard chain away from half filling}
\author{Alexander C. Tiegel}
\affiliation{Institut f\"ur Theoretische Physik, Georg-August-Universit\"at G\"ottingen, 37077 G\"ottingen, Germany}
\author{Thomas Veness}
\affiliation{The Rudolf Peierls Centre for Theoretical Physics, University of Oxford, Oxford, OX1 3NP, United Kingdom}
\author{Piet E. Dargel}
\affiliation{Institut f\"ur Theoretische Physik, Georg-August-Universit\"at G\"ottingen, 37077 G\"ottingen, Germany}
\author{Andreas Honecker}
\affiliation{Laboratoire de Physique Th\'eorique et
Mod\'elisation, CNRS UMR 8089, Universit\'e de Cergy-Pontoise, 95302 Cergy-Pontoise Cedex, France}
\author{Thomas Pruschke}
\thanks{Deceased.}
\affiliation{Institut f\"ur Theoretische Physik, Georg-August-Universit\"at G\"ottingen, 37077 G\"ottingen, Germany}
\author{Ian P. McCulloch}
\affiliation{Centre for Engineered Quantum Systems, School of Physical Sciences, The University of Queensland, Brisbane, Queensland 4072, Australia}
\author{Fabian H. L. Essler}
\affiliation{The Rudolf Peierls Centre for Theoretical Physics, University of Oxford, Oxford, OX1 3NP, United Kingdom}

\date{17.03.2016}                 

\begin{abstract}
We consider the optical conductivity $\sigma_1(\omega)$ in the
metallic phase of the one-dimensional Hubbard model. Our results focus on the
vicinity of half filling and the frequency regime around the optical
gap in the Mott insulating phase. By means of a density-matrix renormalization group
implementation of the correction-vector approach, 
$\sigma_1(\omega)$ is computed for a range of interaction strengths and
dopings. We identify an energy scale $E_{\rm opt}$ above which the
optical conductivity shows a rapid increase. We then use a mobile
impurity model in combination with exact results to determine the
behavior of $\sigma_1(\omega)$ for frequencies just above $E_{\rm  opt}$ which is in agreement with our numerical data. As a main result, we find that this onset behavior is not described by a power law.
\end{abstract}

\maketitle

\newcommand{\RefAbstand}{2mm}
\renewcommand{\d}{\mathrm{d}}
\newcommand{\e}{\mathrm{e}}
\renewcommand{\Im}{\ensuremath{\operatorname{Im}\,}}
\renewcommand{\Re}{\ensuremath{\operatorname{Re}\,}}
\newcommand{\R}{\ensuremath{\mathbb{R}}}
\newcommand{\N}{\ensuremath{\mathbb{N}}}
\newcommand{\Z}{\ensuremath{\mathbb{Z}}}

\newcommand{\ket}[1]{\ensuremath{|#1\rangle}} 
\newcommand{\bra}[1]{\ensuremath{\langle #1|}}
\newcommand{\norm}[1]{\ensuremath{\left\| #1 \right\|}}
\newcommand{\absval}[1]{\ensuremath{\left| #1 \right|}}
\newcommand{\expval}[3]{\ensuremath{ \langle #1 | #2 | #3 \rangle }}
\newcommand{\scprod}[2]{\ensuremath{ \langle #1 | #2 \rangle }}
\newcommand{\average}[1]{\ensuremath{\big<#1 \big>}}
\newcommand{\com}[2]{\ensuremath{ \left[ #1 , #2 \right]_{-} }}
\newcommand{\acom}[2]{\ensuremath{ \left[ #1 , #2 \right]_{+} }}
\newcommand{\GS}{\ensuremath{\mathrm{GS}}}

\newcommand{\cdag}[1]{ \ensuremath{ c_{#1}^\dagger } }
\newcommand{\cann}[1]{ \ensuremath{ c_{#1}^{\phantom{\dagger}} } }

\newcommand{\etal}{\textit{et~al. }}
\newcommand{\pd}{ {\phantom\dagger}}
\newcommand{\up}{\uparrow}
\newcommand{\down}{\downarrow}
\def\nn{\nonumber\\}
\newcommand{\be}{\begin{equation}}
\newcommand{\ee}{\end{equation}}
\newcommand{\beA}{\begin{equation}\begin{aligned}}
\newcommand{\eeA}{\end{aligned}\end{equation}}
\newcommand{\bea}{\begin{eqnarray}}
\newcommand{\eea}{\end{eqnarray}}

\section{Introduction}

The Mott metal-insulator transition is a paradigm for the importance
of electron-electron interactions in correlated many-particle systems. It
occurs in a range of materials and has attracted much attention over 
the last fifty years.\cite{Mott,Gebhardbook} While the mechanism that
drives the transition is well understood, some of the dynamical
properties relating to Mott physics remain to be fully explored. 
A characteristic feature of the Mott phase is the interaction-induced
formation of an excitation gap.\cite{Gebhardbook} This gap is visible in various
dynamical correlation functions such as the real part $\sigma_1(\omega)$ of the optical conductivity
\begin{align}
\sigma_1(\omega) &= - \frac{ {\rm Im}\;\chi^J(\omega)}{\omega}, 
\label{eq:optCondDef1}\\
\chi^J(\omega) &= -\frac{ie^2}{L} \int_0^\infty \d t\ e^{i\omega t}  
\langle GS | [J(t),J(0)]|GS\rangle. \label{eq:optCondDef}
\end{align}
Here $J=\sum_jJ_j$ is the current operator
\begin{equation}
J_j = -it \sum_{\sigma} \left[ c^\dagger_{j,\sigma} c^{\pd}_{j+1,\sigma} - 
c^\dagger_{j+1,\sigma} c^{\pd}_{j,\sigma} \right]. \label{eq:currentOperator}
\end{equation}
The Mott gap disappears upon doping, and an interesting question is
what $\sigma_1(\omega)$ looks like in the metallic phase close
to the Mott transition. Here we investigate this issue in one spatial
dimension for the archetypal example of the Mott transition, the
Hubbard model\cite{book} 
\begin{align}
	H &= -t \sum_{j,\sigma} \left[ c^\dagger_{j+1,\sigma} c^{\pd}_{j,\sigma}  
	+ c^\dagger_{j,\sigma}c^{\pd}_{j+1,\sigma}  \right]
	+ U\sum_j n^{\pd}_{j,\up} n^{\pd}_{j,\down} \nonumber \\
	&-\mu \sum_j \left[ n^\pd_{j,\up} + n^\pd_{j,\down} \right]. \label{eq: Hamiltonian}
\end{align}
Here, $c^{\phantom\dagger}_{j,\sigma}$ annihilates a fermion with spin 
$\sigma=\, \up,\down$ at site $j$, 
$n^{\phantom\dagger}_{j,\sigma}=
c^\dagger_{j,\sigma}c^{\phantom\dagger}_{j,\sigma}$ 
is the number operator, $t$ is the hopping parameter which is set to $t=1$ in our calculations, $\mu$ is the chemical
potential, and $U\geq 0$ is the strength of the on-site repulsion. 

\begin{figure}[ht]
\centering
\includegraphics[width=0.9\columnwidth]{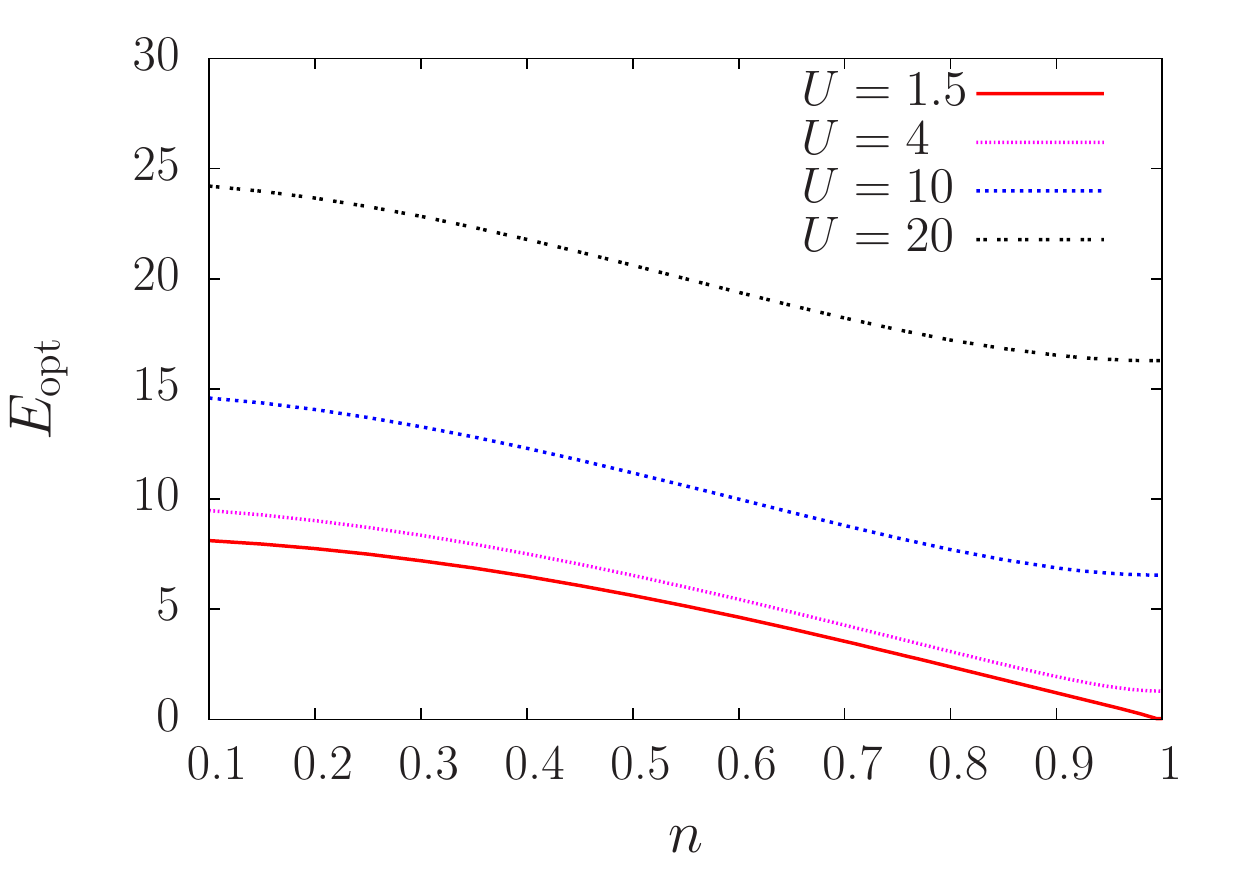}
\caption{(Color online) Bethe ansatz results for theoretical ``pseudogap'' onset value\cite{PhysRevLett.84.4673,inprep} $E_{\rm
    opt}$ as a function of $U$ and $n$. } 
  \label{fig:eopt}
\end{figure}

At zero temperature and half filling the optical conductivity has been
comprehensively analyzed by both analytic and numerical methods:\cite{PhysRevLett.85.3910, PhysRevB.66.045114, PhysRevLett.86.680,
  PhysRevB.64.125119} the system is insulating and there is an 
optical gap\cite{Ovchi_1970} at $\omega = 2\Delta$, i.e., twice the Mott-Hubbard gap, below which 
the optical conductivity vanishes. Immediately above this gap,
$\sigma_1(\omega)$ exhibits a square-root increase.  
In contrast, much less is known regarding the optical conductivity
away from half filling. In the thermodynamic limit, the optical
conductivity consists of a delta peak at zero frequency, the Drude
peak, and a so-called regular or incoherent part 
\begin{equation}
\sigma_1(\omega\geq 0) = D \, \delta(\omega) + \sigma^{\rm reg}(\omega).
\end{equation}
The low-frequency behavior has been studied using methods based on
Luttinger liquid theory,\cite{PhysRevB.44.2905,PhysRevB.46.9325, Giamarchi1997975}
which predict a universal $\omega^3$ behavior of $\sigma^{\rm reg}(\omega)$ 
at $0<\omega \ll t$. 
Moreover, in the case of one doped hole at strong coupling ($U \gg 1$), an $\omega^{3/2}$ dependence at small frequency and spectral weight in the region $0 < \omega < 4t$ has been reported.\cite{PhysRevB.48.10595}. 
However, it is clear on general
grounds that at low dopings, i.e.\,$1-n\ll 1$, only a minute fraction of the
total spectral weight in $\sigma_1(\omega)$ will be associated
with features at frequencies below the optical gap $2\Delta$ at half filling. 
One expects there to be a characteristic ``pseudogap''
energy scale $E_{\rm opt}$ above which $\sigma_1(\omega)$ will
increase and exhibit a similar behavior to the one seen at half filling. 
The low-intensity features below $E_{\rm opt}$ involve only excitations comprising of holon-antiholon pairs.
The scale $E_{\rm opt}$ has been identified in a work by Carmelo\,\etal\cite{PhysRevLett.84.4673} and is obtained from the Bethe
ansatz solution of the one-dimensional Hubbard model.\cite{book} In
Fig.~\ref{fig:eopt} we present results for $E_{\rm opt}$ as a function of the
band filling for several values of $U$.

In Ref.~\onlinecite{PhysRevLett.84.4673} it was conjectured that the
optical conductivity increases in a power-law fashion above $E_{\rm opt}$
\begin{align}
\sigma_1(\omega) \sim \left( \omega - E_{\rm opt} \right)^\zeta \, 
   \Theta(\omega-E_{\rm opt})\ .
   \label{eq:carmeloPrediction}
\end{align}
In the following we investigate the behavior of the optical
conductivity for small dopings, paying particular attention to its
behavior above the pseudogap. Our analysis is based on a combination
of density-matrix renormalization group (DMRG) computations\cite{PhysRevLett.69.2863} and results
obtained by employing a mobile impurity model description 
\cite{SIG,PS,FHLE,Pereira,Seabra,EsslerPereira,inprep} augmented by exact Bethe
ansatz calculations. 

The paper is organized as follows. We briefly review the DMRG-based correction-vector approach in Sec.~\ref{sec: numerical_method} and present our numerical results in Sec.~\ref{sec: absorption_band}. The frequency dependence of $\sigma_1(\omega)$ above $E_{\rm opt}$ is also determined by means of a mobile impurity model (MIM) in combination with exact results in Sec.~\ref{sec: MIM}. Section~\ref{sec: onset_behavior} provides a comparison of our DMRG and MIM calculations, which shows that the onset behavior directly above $E_{\rm opt}$ is not described by a power law. Finally, our conclusions are summarized in Sec.~\ref{sec: conclusion}.

\section{Numerical method} \label{sec: numerical_method}

We use a matrix product state (MPS)\cite{Ulrich201196,McCulloch_2007} 
implementation of the correction-vector approach,\cite{soos:1067} which is an 
extension of the DMRG to compute spectral functions. There exist several 
 variants of this correction-vector approach\cite{PhysRevB.54.7598,PhysRevB.60.335,PhysRevB.80.165117} such as DDMRG.\cite{PhysRevB.66.045114} We can recast Eq.\ \eqref{eq:optCondDef1} as
\begin{align}
	\sigma_1 (\omega  > 0) = - \lim_{\eta\to0^+} \frac{e^2}{\omega L} 
	\Im G_{J}(\omega  > 0,\eta),
\end{align}
where
\begin{align}
G_{J} (\omega,\eta) = \langle GS | J^\dagger\, 
\frac{1}{\omega+i\eta - (H-E_{GS})} \, J | GS \rangle \label{eq: dyn_corr_func}.
\end{align}
Here $E_{GS}$ is the ground-state energy. 
The correction vector is defined by
\begin{align}
    \ket{\psi_J (\omega ,\eta)} =  
    \frac{1}{E_{GS}+\omega + i \eta - H } J | GS \rangle 
    \label{eq: corr_vec},
\end{align}
and can be obtained as the solution $\ket{\psi}$ of the linear system
\begin{align}
	(E_{GS} + \omega + i \eta - H) \ket{\psi} =  J | GS \rangle. 
    \label{eq: linear_system}
\end{align}
Here the basic idea is to variationally determine the correction vector 
associated with $G_{J} (\omega,\eta)$ at the frequency of interest within 
the ansatz class of MPS. We solve this set of equations directly by 
local updates of the MPS $\ket{\psi}$ (see Ref.\,\,\onlinecite{ediss13933} for 
details). Sweeping through the chain in a DMRG-like fashion until convergence 
is reached, a local non-Hermitian system of equations is solved at each site by 
the generalized minimal residual (GMRES) method.\cite{Saad_2003} The 
dynamical correlation function can be evaluated as the overlap 
$G_{J} (\omega, \eta) = \langle GS |J^\dagger |\psi_J (\omega , \eta) \rangle $. 
Note that the correction vector needs to be computed separately for each
frequency $\omega$. Importantly, the method gives intrinsically broadened 
results with a Lorentzian line shape of width $\eta>0$, which is crucial for 
Eq.~\eqref{eq: linear_system} to be well conditioned. 
The correction-vector calculations are performed 
for chains of up to $L = 84$ sites and open boundary conditions (OBCs).
Finite-size effects cause the spectral weight of the Drude peak to be redistributed to finite frequencies above the lowest energy scale $\sim 1/L$.\cite{PhysRevB.44.6909}
By considering sufficiently large $U$, these effects are well separated 
from the onset at the edge of the ``pseudogap''. 
To obtain accurate results, we exploit the SU(2) symmetry\cite{McCulloch_2007} of the Hamiltonian \eqref{eq: Hamiltonian} and keep $m=1300$ DMRG states for ground-state calculations.
For the dynamics, $m=500$ states were retained for the correction-vector approach at a filling of $n=1$ and
at $n<1$, $m=600$.

\section{Results for the absorption band} \label{sec: absorption_band}
\begin{figure}[b]
  \centering
  \includegraphics[width=0.9\columnwidth]{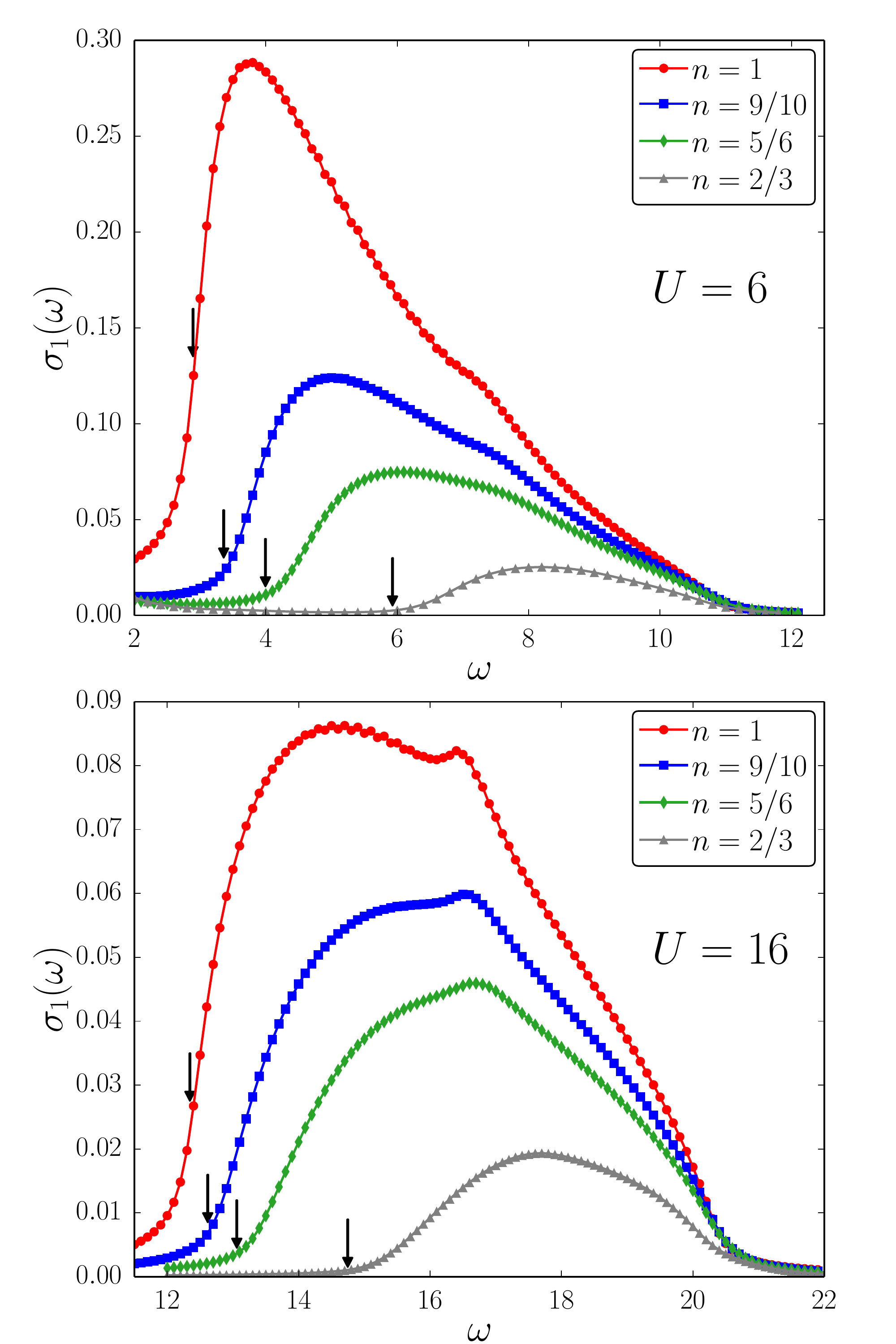}
  \caption{(Color online) DMRG results for the regular part of the optical conductivity  show a well-defined absorption band for various fillings $n$ and a Lorentzian broadening of $\eta=0.2$. The data are obtained for a chain of $L=60$ and OBCs. (Upper panel) $U=6$.  
  (Lower panel) $U=16$. The arrows mark the results for $E_{\rm opt}$ determined by Bethe ansatz. Note that the $\omega$ axis starts at different 
  frequencies in both panels.}
  \label{fig:cv_L60_overview}
\end{figure}

A well-defined absorption band above $E_{\rm opt}$ is only observed for sufficiently large values of the repulsion $U$. Moreover, far from half filling, e.g.\,at quarter filling ($n=1/2$), almost all of the intensity is contained in the Drude peak.\cite{PhysRevLett.64.2831} Therefore, the DMRG results for $\sigma_1(\omega)$ in Fig.~\ref{fig:cv_L60_overview} are obtained for $U=6$ and 16 and a filling factor not smaller than $n=2/3$.
Finite-size and boundary condition effects are dominated by 
the intrinsic broadening introduced due to $\eta=0.2$.
Only the $n=2/3$ curve in the upper panel of 
Fig.~\ref{fig:cv_L60_overview} displays a slight increase towards small 
frequencies. This increase is mainly a consequence of the Drude peak appearing at finite frequencies for OBCs\cite{PhysRevB.44.6909} and the growing Drude weight for a fixed value of $U$ with increasing doping. Moreover, it is observed that for a given $U$ the integrated spectral weight below the regular part decreases with increasing doping. This is in qualitative agreement with exact results for the relative weight of 
Drude peak with respect to the total intensity for an infinite system.\cite{PhysRevLett.64.2831}
In the thermodynamic limit, the Drude peak vanishes at half filling; 
this transfer of spectral weight to finite frequencies as a function of $n$ is 
very sharp for small and extremely large $U$ and can be understood in terms of
umklapp processes.\cite{PhysRevLett.64.2831}
The two spectra at half filling ($n=1$) shown in
Fig.~\ref{fig:cv_L60_overview} are in agreement with the $U$
dependence of the Mott-Hubbard gap $\Delta$.\cite{book} With decreasing filling $n$, the frequency 
at which $\sigma_1(\omega)$ becomes sizable for a given $U$ increases compared to
$2\Delta$. The rapid increase of the DMRG results above this frequency agrees well with existing results for $E_{\rm opt}$,\cite{PhysRevLett.84.4673} which are marked by arrows in Fig.~\ref{fig:cv_L60_overview}.  The
broadened spectra also suggest that the onset at the lower threshold becomes softer for decreasing filling 
and the upper band edge does not vary significantly for different fillings. 
This softening can be understood in terms of the mobile impurity
approach discussed below.
The results in Fig.~\ref{fig:cv_L60_overview} confirm the
expectation that the optical spectra become more symmetric for higher
values of $U$.\cite{PhysRevLett.85.3910, PhysRevLett.84.4673} The
small peak in the middle of the absorption band is very similar to the
one previously observed at half filling,\cite{PhysRevLett.85.3910}
and has its origin in the large density of states for excitations
between parallel bands. Its existence is evident for $U=16$ and it can
still be observed as a weak feature for $U=6$. For $U=16$ the small
peak is found to persist at least down to $n=5/6$.

\begin{figure}[t]
  \centering
  \includegraphics[width=0.9\columnwidth]{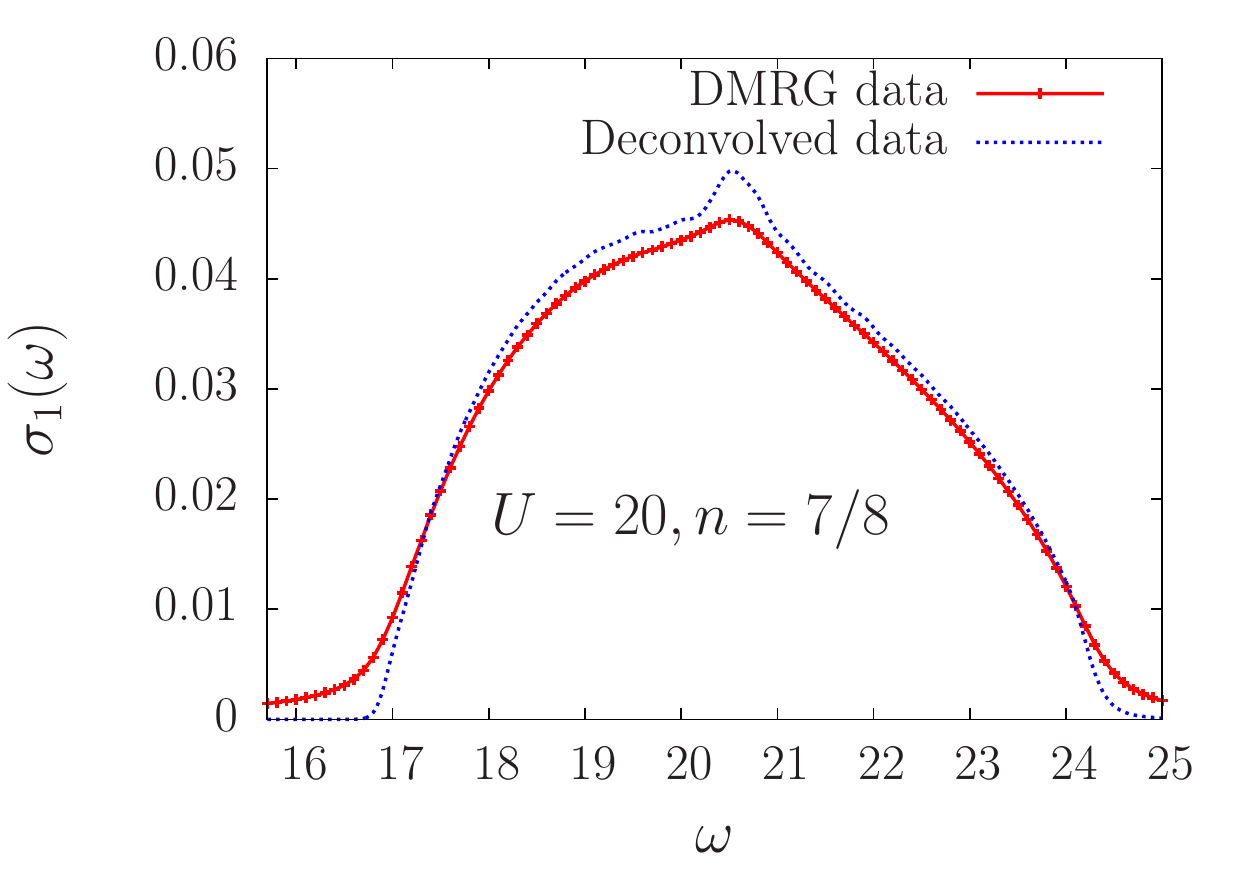}
  \includegraphics[width=0.9\columnwidth]{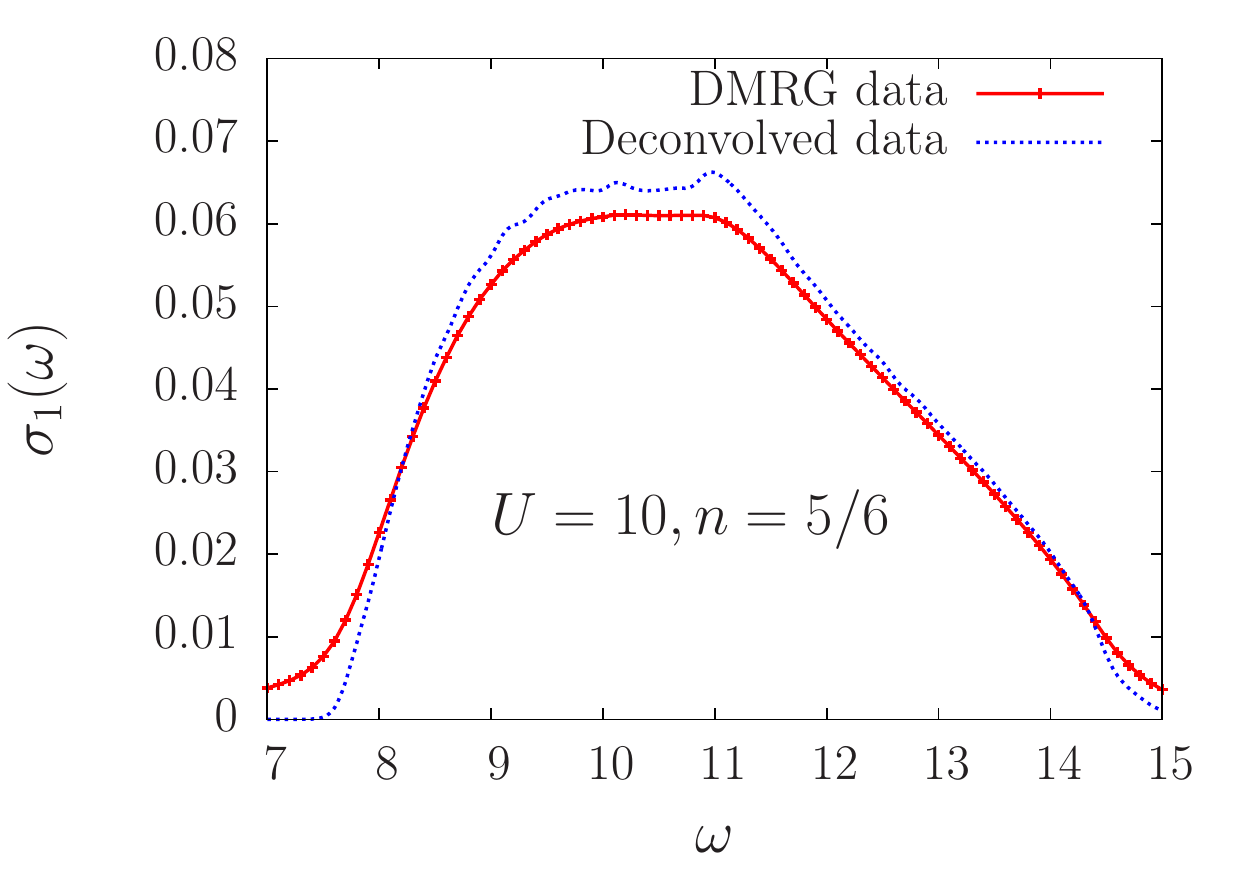}
  \caption{(Color online) DMRG results with an intrinsic Lorentzian broadening of $\eta=0.2$ are compared to the corresponding deconvolved data. (Upper panel) $U=20$, $n=7/8$, and $L=80$. (Lower panel)  $U=10$, $n=5/6$, and $L=84$.}
  \label{fig:deconv}
\end{figure}

In order to compare our DMRG results to the prediction of the mobile impurity model (MIM) presented in the next section, it is necessary to remove the intrinsic
Lorentzian broadening of the DMRG data. This a numerically ill-conditioned
problem, but in practice the following procedure was found  to work reliably. The initial correction-vector results are obtained on a grid of frequencies separated
by $\Delta \omega = 0.1$. We use rational functions to both
interpolate and extrapolate this data.\cite{numRec} The resulting
continuous function is then deconvolved using the Richardson-Lucy
algorithm.\cite{RL1,RL2}
Comparisons of the inherently broadened DMRG results and the deconvolved data are
presented in Fig.~\ref{fig:deconv}, where the onset behavior
is smooth, but small artifacts can be seen at higher frequencies.

\section{Mobile Impurity Model (MIM)} \label{sec: MIM}

\begin{figure}[t]
  \centering
  \includegraphics[width=0.9\columnwidth]{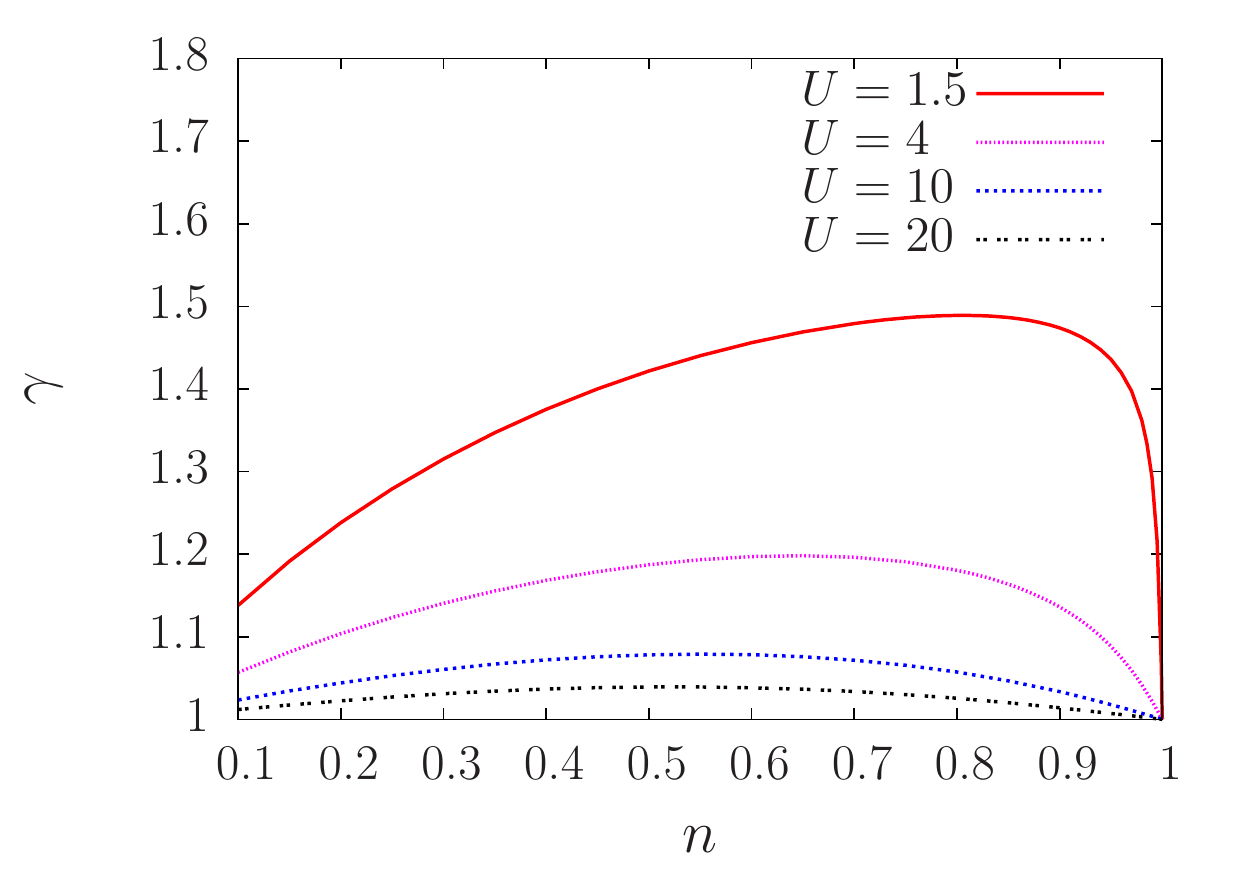}
  \caption{(Color online) Parameter $\gamma$ as a function of $U$ and $n$. }
  \label{fig:gamma}
\end{figure}

While the low-energy sector of the Hubbard model is described by a
spin-charge separated Luttinger liquid (LL), the calculation of finite-frequency properties requires a careful treatment of
perturbations. Perturbation theory in some of these
irrelevant operators exhibits infrared singularities, which lead to
strong deviations from LL behavior. Crucially, in the vicinity of
thresholds for simple excitations the problem can be mapped to
that of a high-energy mobile impurity coupled to a LL.\cite{SIG,PS} The parameters of this MIM can be completely
determined by using exact results obtained in the framework of the
Bethe ansatz solution.\cite{FHLE} The appropriate model for the
optical conductivity at frequencies just above $E_{\rm opt}$ can be
cast in the form $H= \int dx[{\cal H}_{\rm
    LL}+{\cal H}_{\rm imp}+{\cal H}_{\rm int}]$, where 
\cite{EsslerPereira,inprep} 
\bea
{\cal H}_{\rm LL}&=&
\sum_{\alpha=c,s}\frac{v_\alpha}{16\pi}\left[\frac{1}{2K_\alpha}
\big(\partial_x\Phi_\alpha^*\big)^2
+2K_\alpha\big(\partial_x\Theta_\alpha^*\big)^2\right], \nn
{\cal H}_{\rm  imp}&=&
B^\dagger(x)\left[\varepsilon(0)-\frac{1}{2}\varepsilon''(0)\partial_x^2\right]B(x)\ ,\nn 
{\cal H}_{\rm int}&=&
B^\dagger(x)B(x)\left[f_\alpha\partial_x\varphi^*_\alpha(x)+
\bar{f}_\alpha\partial_x\bar{\varphi}^*_\alpha(x)\right].
\label{MIM}
\eea
Here the Luttinger liquid part ${\cal H}_{\rm LL}$ describes the
low-energy spin and charge collective modes, whereas ${\cal H}_{\rm
  imp}$ is the Hamiltonian of a high-energy ``impurity''
with quadratically decreasing dispersion $\varepsilon(p)$ around zero
momentum. Finally ${\cal H}_{\rm int}$ describes the interaction of
the impurity with the low-energy degrees of freedom. The parameters 
$v_{c,s}$, $K_{c,s}$, $f_{c,s}$, $\bar f_{c,s}$, and $\varepsilon(q)$
in \eqref{MIM} can be determined from the Bethe ansatz 
solution.\cite{inprep} The physical
content of the model \eqref{MIM} is as follows. Excitations at
frequencies just above $E_{\rm opt}$ consist of a single high-energy
bound state ($k$-$\Lambda$ string\cite{book}) and a number of
low-energy excitations.\cite{PhysRevLett.84.4673} Assuming the bound
state to be a point-like object and retaining only the most relevant
interactions in ${\cal H}_{\rm int}$ then leads to the model
(\ref{MIM}). The current operator \eqref{eq:currentOperator} can be
projected on the MIM degrees of freedom\cite{inprep}
\be
J_j\rightarrow 
\left(\partial_x B^\dagger(x)\right) e^{-i
  \Theta^*_c(x)/\sqrt{2}} \sin
\left(\frac{\Phi_s^*}{2\sqrt{2}}\right)+\ldots 
\label{currentproj}
\ee
The calculation of the current-current correlation function, and thus
the optical conductivity in the framework of the MIM \eqref{MIM}, then
proceeds along standard lines \cite{SIG} and results in an expression
of the form
\begin{widetext}
\begin{eqnarray}
\sigma_1(\omega\approx E_{\rm opt}) &\sim& \frac{C}{\omega}
\int_{-\Lambda}^\Lambda \d p  \Big\{
\frac{\gamma_c^2}{K_c} \left(
\left( 1 + \gamma\right) 
\left[\widetilde{G}^{c}_{\gamma+2,\gamma}\big(\omega,p\big)  
+ \widetilde{G}^{c}_{\gamma,\gamma+2}\big(\omega,p\big)\right] 
- 2\gamma\widetilde{G}^{c}_{\gamma+1,\gamma+1}\big(\omega,p\big) 
\right) \nn
&&\qquad+\sqrt{\frac{4\gamma}{K_c}} \gamma_c p \left[ 
\widetilde{G}^c_{\gamma+1,\gamma}\big(\omega,p\big) 
-\widetilde{G}^c_{\gamma,\gamma+1}\big(\omega,p\big) \right]
+ p^2 \widetilde{G}^{c}_{\gamma,\gamma}\big(\omega,p\big)+\gamma_s^2
\widetilde{G}^{s}_{\gamma}\big(\omega,p\big) 
\Big\}, \label{eq:klOnsetForm}
\end{eqnarray}
\end{widetext}
where
\be
\widetilde{G}^{c}_{\gamma,\delta}(\omega,p) = 
\frac{(2\pi)^2\Theta(\omega_c(|p|))
(\omega_c(-p))^{\gamma-1} (\omega_c(p))^{\delta-1}
}
{\Gamma(\gamma)\Gamma(\delta)(2v_c)^{\gamma+\delta-1}},
\label{eq:Gtildec}
\ee
\bea
\widetilde{G}^{s}_{\gamma}(\omega,p) &=&
\frac{(2\pi)^2(\omega_s(p))^{2\gamma-1}}{\Gamma^2\left( \gamma \right)
(v_c^2-v_s^2)^{\gamma}} 
\Theta(\omega_s(p))\int_0^1 \d s  s^{\gamma-1}\nn
&\times&(1-s)^{\gamma-1}
\left[\frac{2v_c(\omega-v_sp)}{v_c^2-v_s^2} s- \frac{\omega_c(p)}{v_c-v_s}  \right]
\nn
&\times&
\Theta \Big(
\frac{2v_c\omega_s(p)}{v_c^2-v_s^2}s
 - \frac{\omega_c(p)}{v_c-v_s}  \Big)+p\rightarrow -p .
\end{eqnarray}
Here we have defined $\omega_\alpha(p)=\omega-\varepsilon(p)-v_\alpha p$, 
and $\Lambda$ is a cutoff. The parameter $\gamma$ is shown in
Fig.~\ref{fig:gamma} as a function of band filling for several values
of $U$. 
The result (\ref{eq:klOnsetForm}) applies in an a priori unknown frequency
window above $E_{\rm opt}$. This energy window shrinks to zero as we
approach half filling $n\to1$, and the behavior of
(\ref{eq:klOnsetForm}) is in fact very different from the square-root
increase seen at half filling.

\section{Behavior of $\sigma_1(\omega)$ above the
  crossover scale $E_{\rm opt}$} \label{sec: onset_behavior}

  Focusing on frequencies in the vicinity of $E_{\rm opt}$ in
Fig.~\ref{fig:deconv}, we observe that the deconvolved DMRG data
exhibits a smooth and slow increase. This behavior can be directly 
compared to the results obtained from the MIM. 
\begin{figure}[h]
  \centering
  \includegraphics[width=0.9\columnwidth]{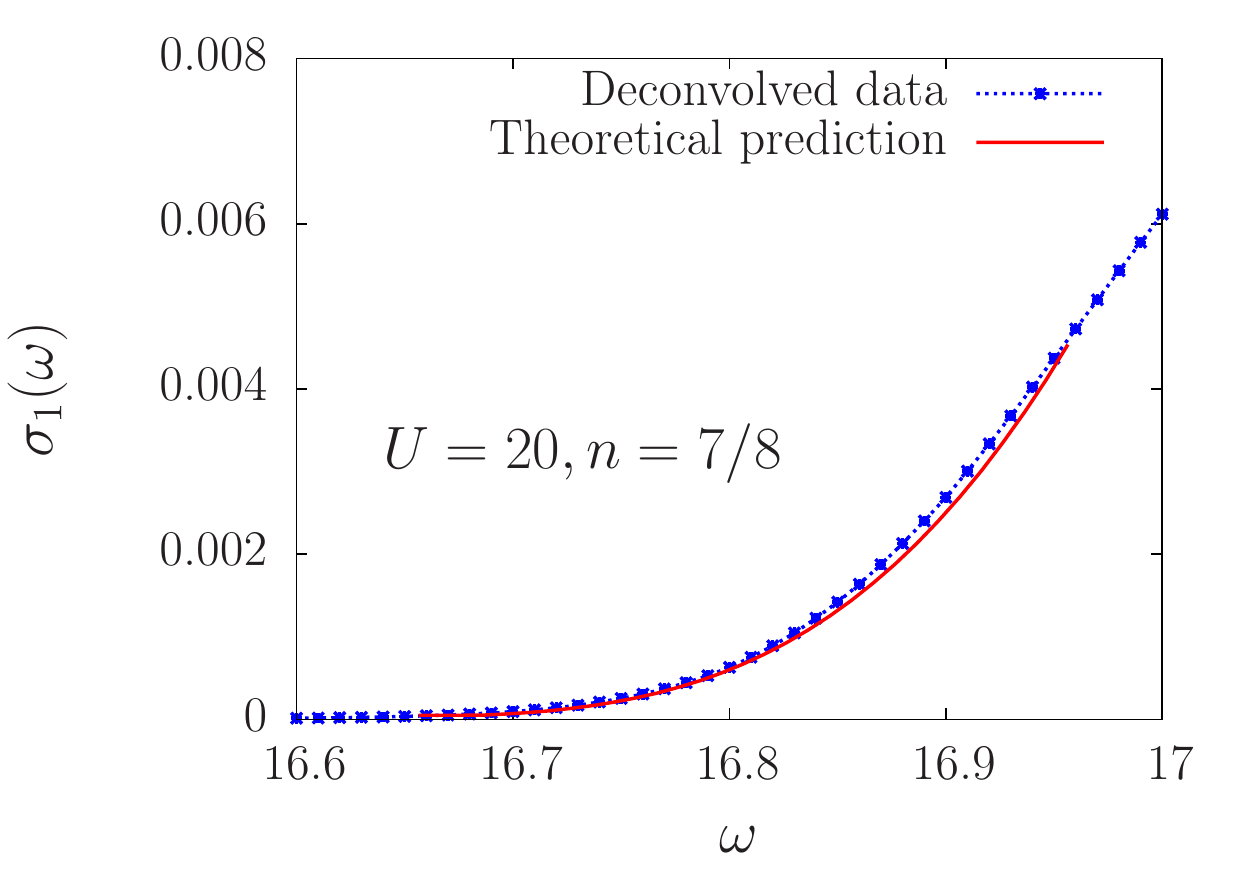}
  \includegraphics[width=0.9\columnwidth]{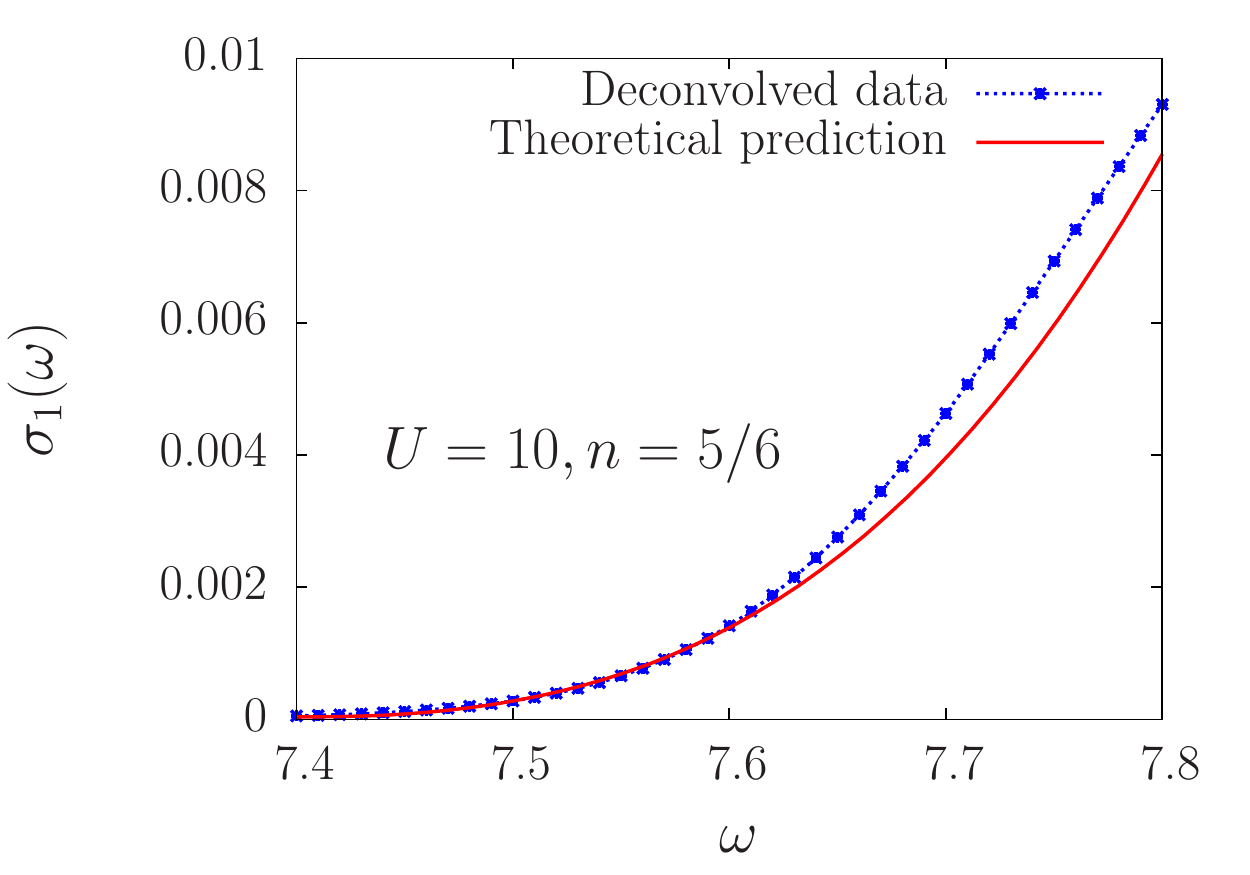}
  \caption{(Color online) Comparison of theoretical predictions from a mobile impurity model 
  with numerical results from DMRG. }
  \label{fig:comp}
\end{figure}
In the latter we adjust the overall amplitude $C$ allowing for a small,
constant contribution attributed to excitations involving only holons
and antiholons which give rise to the Drude peak at zero frequency,
but are expected to make up only a small fraction of the spectral
weight at $\omega\approx E_{\rm opt}$. We furthermore adapt the 
cut-off $\Lambda$, although the results depend only weakly on it.
The comparison in Fig.~\ref{fig:comp} shows that the MIM results are
consistent with the deconvolved DMRG data. Moreover, the increase in
$\sigma_1(\omega)$ above $E_{\rm opt}$ is not described by a
power law. On a technical level this can be traced back to the fact
that the mobile impurity sits at a maximum of its dispersion relation.
The results obtained by means of the MIM are very different from
the power-law increase \eqref{eq:carmeloPrediction} predicted in
Ref.~\onlinecite{PhysRevLett.84.4673}. In particular the exponent $\zeta$
predicted in this previous work becomes less than one for $U>4$, which is not
consistent with our deconvolved DMRG data.

\section{Conclusions} \label{sec: conclusion}
We have studied the real part $\sigma_1(\omega)$ of the
zero-temperature optical conductivity in the one-dimensional Hubbard
model in the metallic phase close to half filling. At half filling
$n=1$, it is known that $\sigma_1(\omega)$ vanishes below twice the
Mott gap, and then increases in a characteristic square-root fashion.
\cite{PhysRevLett.85.3910} Doping away from half filling induces
a Drude peak at zero frequency, the weight of which scales with $1-n$.
Here we have focused on frequency scales close to the optical gap at
half filling, and investigated how $\sigma_1(\omega)$ gets modified
upon doping holes into the system. In our DMRG calculations, we have observed a rapid increase above a crossover scale $E_{\rm opt}$, and analyzed this behavior in the framework of a mobile impurity model. The results obtained by this 
method were found to be in agreement with our DMRG data. Therefore, the increase of $\sigma_1(\omega)$ for frequencies above the pseudogap $E_{\rm opt}$, in which only small-amplitude excitations consisting  of holon-antiholon pairs are present, is not described by a power law.

\textit{Note added:} We regret to announce that one of our coauthors, Prof.\ Thomas Pruschke, passed away shortly after the submission of this article. We would like to express our gratitude for his unflagging support as a colleague and his incisive contributions as a physicist. 

\acknowledgments
We thank Imke Schneider for helpful discussions and collaboration in the early stages of this work. We acknowledge the support by the Helmholtz Association via the Virtual Institute ``New states of matter and their excitations'' (Project No. VH-VI-521). This work was supported by the EPSRC under Grants No.  EP/I032487/1 and No. EP/J014885/1 (FHLE and TV). IPM acknowledges the support from the Australian Research Council Centre of Excellence for Engineered Quantum Systems, CE110001013, and the srtFuture Fellowships scheme, FT100100515.

\end{document}